# THE LIBERALITIES AND TYRANNIES OF ICTs FOR VULNERABLE MIGRANTS: THE STATUS QUO, GAPS AND DIRECTIONS

Yidnekachew Redda Haile, Royal Holloway University of London, Yidu.Haile.2019@live.rhul.ac.uk / yidnekr@yahoo.com

**Abstract:** Information and communication technologies (ICTs) have increasingly become vital for people on the move including the nearly 80 million displaced due to conflict, violence, and human right violations globally. However, existing research on ICTs and migrants, which almost entirely focused on migrants' ICT use 'en route' or within developed economies principally in the perspectives of researchers from these regions, is very fragmented posing a difficulty in understanding the key objects of research. Moreover, ICTs are often celebrated as liberating and exploitable at migrants' rational discretion even though they are 'double-edged swords' with significant risks, burdens, pressures and inequality challenges particularly for vulnerable migrants including those forcefully displaced and trafficked. Towards addressing these limitations and illuminating future directions, this paper, first, scrutinises the existing research vis-à-vis ICTs' liberating and authoritarian role particularly for vulnerable migrants whereby explicating key issues in the research domain. Second, it identifies key gaps and opportunities for future research. Using a tailored methodology, broad literature relating to ICTs and migration/development published in the period 1990-2020 was surveyed resulting in 157 selected publications which were critically appraised vis-à-vis the key themes, major technologies dealt with, and methodologies and theories/concepts adopted. Furthermore, key insights, trends, gaps, and future research opportunities pertaining to both the existing and missing objects of research in ICTs and migration/development are spotlighted.

## 1. INTRODUCTION

Despite a 27 per cent slowdown precipitated by COVID-19, as of 2020, over 281 million people including around 80 million displaced due to conflict, violence and human right violations—of which around 34 million are border crossing refugees and "asylum seekers"—are on the move (UN DESA, 2020; UNHCR, 2020). Meanwhile, information and communication technologies (ICTs) or 'digital technologies' are increasingly becoming an integral part of the dynamics. ICTs, primarily mobile phones and the internet, are significantly affecting contemporary human mobility and migration, labour and resource flows, globalised economy and information sharing. As such, these technologies have become essential components in the lives of people on the move especially vulnerable migrants such as the displaced, trafficked or exploited, as well as their families and networks in all stages of migration—pre-migration, migratory journeys, (re)settlement/"integration" and return migration (Hamel, 2009; Massey, 2009; Surak, 2017). Descriptions such as the "connected migrants" (Diminescu, 2008), "digital migrants/diaspora/refugee" (Everett, 2009;

Jumbert, 2018), "mediatized migrant" (Hepp et al., 2012) and "connected presence" (Licoppe, 2004) all reflect the entwinement of these technologies with migrants. Advances in these technologies have increasingly complicated people's movements and connections to more circulatory, complex, dynamic, fluid and unpredictable patterns (Castells, 2011). The distinction between mobility and migration as well as other related concepts such as migrant, foreigner, immigrant, diaspora, nomad, refugee, asylum migrant, (internally) displaced person (IDP), irregular migrant, economic migrant and even sedentary (Douglas et al., 2019) has become more fuzzy and problematic (Crawley & Skleparis, 2018).

Recently, interest in this entwinement of ICTs and migration/development has enormously increased among academics, civil society organisations, international agencies, and governments. Although explicit research on migrants' use of new ICTs can be traced back to at least the 1990s, the area still remains young demanding more research (Andersson, 2019). This could be attributed to the ever-changing complexity and dynamism in the way people move, interact, and connect amidst the rapid shifts and changes in technology and innovation. Existing research in ICTs and migrants—which is almost entirely focused on migrants' use of technology 'en route' or within countries with developed economies ("Global North") principally in the perspectives of researchers from these regions—remains very fragmented. Furthermore, the increasingly used phrases such as "migrant tech" and "refugee tech" are so rosy celebrating ICTs as liberating and utilisable at migrants' discretion and rational choice despite the reality that they are 'double-edged swords' with significant risks, burdens, pressures and inequality challenges particularly for vulnerable migrants.

This paper attempts not least a twofold contribution to ICTs and migrants' research towards addressing the aforementioned limitations. First, it scrutinises the status quo of such research vis-à-vis the liberating and authoritarian power of ICTs particularly for vulnerable migrants whereby explicating key research objects for future theorisation. Second, it identifies key gaps and opportunities for future research. To this end, broad literature relating to ICTs and migration/development published between 1990 and 2020 was surveyed producing 157 selected publications—not all are cited in-text for brevity and space/word limitation but they are outlined in Table 1. The selected literature was critically appraised vis-à-vis the key themes, major technologies dealt with, and methodologies and theories/concepts adopted. Key insights and trends were also gleaned.

Traditionally, mobility has been understood as a transitory movement of people (also capital, goods and services) in the course of their everyday life including livelihood mobility (Benz, 2014; Steel et al., 2017) and residential mobility (Cooke & Shuttleworth, 2018); and migration as a permanent change of location often between countries or urban-rural areas (Gmelch, 1980). Various parties adopt their version of migration-related concepts for their purpose (such as legal, administrative, academic and statistical) informed by spatial, temporal, methodological, legal, cultural and political factors (IOM, 2007; 2019). Attributes such as places of birth and residence, legal rights and citizenship and length of stay influence the way these concepts are defined and used. As Sironi et al. (2019) affirm, such definitions are sometimes 'inclusivist' and other times 'residualist' (exclude some parties such as those who flee war or persecution). For example, the UN agency, IOM defines migration as:

> A process of moving, either across an international border, or within a State. It is a population movement, encompassing any kind of movement of people, whatever its length, composition and causes; it includes migration of refugees, displaced persons, uprooted people, and economic migrants (IOM, 2004, p. 41).

Similarly, according to the 1951 Geneva convention, 'refugee' is defined as:

> A person who, owing to a well-founded fear of persecution for reasons of race, religion, nationality, membership of a particular social group or political opinion, is outside the country of his [sic] nationality and is unable or, owing to such fear, is unwilling to avail himself of the protection of that country; or who, not having a nationality and being outside the country of his former habitual residence as a result of such events, is unable or, owing to such fear, is unwilling to return to it (IOM, 2021).

This paper primarily focuses on vulnerable migrants such as those displaced, trafficked, and exploited notwithstanding the problems associated with labelling and attempting a clear-cut distinction between different "types" of migrants noted earlier. The author believes that this work is the first overarching rigorous analysis of ICTs and migrants; notwithstanding some related work—for example, Brown et al. (2019) reviewed 35 papers on refugees' ICT use; Collin et al. (2015) identified three research areas in migrants' ICT use—to prepare for migration, to maintain link and variation in use; Chonka & Haile (2020) reviewed literature relating to the role of ICTs in mobility and migration pertinent to the Horn of Africa; and Andersson (2019) overviewed "diaspora" and new media use. This analysis has the potential to substantially inform future research and policy; and shed light on the importance of ICTs, mobility, and migration for ICTs for Development (ICT4D) research.

Following this introduction, Section 2 presents the methodology devised and used. Sections 3 and 4 respectively elucidate empirical evidence, methods, trends, and insights on the liberating and constraining power of ICTs for migrants based on the existing literature; followed by Section 5 which denotes the gaps and future research opportunities. The paper ends with a short concluding remark in Section 6.

## 2. METHODOLOGY

This section presents a short discussion of the methodology adopted (See 'Appendix I' for the detailed methodology). To add rigour, a literature review strategy that draws mainly on Fink (2019), Gall et al. (1996), Geldof et al. (2011), Iden & Eikebrokk (2013), Labaree (2020), Okoli (2015), and Okoli & Schabram (2010) was devised to identify, collect, evaluate, select and analyse the literature. The selected literature was pooled from diverse areas related to ICTs, migration and development. These include media and communication, information systems, information and communication technologies, information and communication technologies for development (ICT4D), migration/mobility, diaspora, transnationalism, development studies, law, geography, economics, sociology, anthropology and history.

A mix of material accessed through searching catalogues, electronic databases, academic journals, conferences, and search engines principally Google (Scholar) was analysed. While the review mostly focused on peer-reviewed published academic work (mainly from 20 journals), other relevant grey literature such as practitioner reports, policy documents, working papers, theses as well as books were also considered. While this review may be considered as primarily integrative with historical, theoretical, methodological, systematic and argumentative features (Labaree, 2020; Okoli, 2015), I purposely avoided using the commonly used adjective "systematic" to describe it. This is due to the inherent limitations of the so called "systematic" literature review particularly in the context of multidisciplinary social science reviews conducted by limited reviewers within a constrained timeframe which may lead to exacerbated generalisations. While this analysis was conducted by a single reviewer as part of doctoral research, to minimize 'reviewer bias', a stringent review procedure was followed, extended time (around 18 months) was allowed, and the work has gone through several iterations based on feedback from four experienced academics. However, while the approach adopted here resonates with the "systematic" literature review techniques, I, by no means, claim the review to be exhaustive to generate definitive generalisations and bias free.

Generic themes and topics were identified using a hybrid of manual and software-assisted automated inductive coding. The first set of themes and topics were manually identified by reviewing 117 papers selected from the initial relevant literature retrieved by searching databases using the combination of keywords that represent all key concepts, i.e., digital technology/ICT, migration, and inequality/development. However, as the keyword search strategy was expanded and the themes started to recur, inspected automatic coding was considered for the additional literature. Consequently, all of the selected publications including those manually thematised were imported to 'NVivo 12 plus' content management and analysis software package and coded inductively. After a careful manual review and validation of the automated coding to address limitations associated with algorithmic evaluation, themes emerged from the manual analysis and from the automated coding were compared, collated, merged, and refined resulting the key themes. The main themes emerged are organised and examined vis-à-vis ICTs' liberating and tyrannical role for migrants as follows.

## 3.    THE LIBERALITIES OF ICTS FOR MIGRANTS

In most literature, ICTs are viewed as liberating for migrants with little or no consideration to their negative aspects. This section scrutinises such literature with respect to four themes which emerged in the analysis; the facilitating role of ICTs for: "diasporic", transnational, and familial connections; planning and execution of migration; migrants' "integration"; and migration-related livelihoods.

### 3.1.   ICTs for facilitating "diasporic", transnational, and familial connections

While concentrated in the contexts and perspectives of the "Global North", a relatively large and relatively established body of literature explores how the "diaspora" or transnational community often settled in "Western" countries are connected. It underlines how ICTs have provided a platform for migrants to stay connected with their family and community whereby maintaining their culture and identities and transnational ties; although ICTs are mostly implicit (Hall, 1990; Levitt and Jaworsky, 2007). Such literature examines how ICTs enable the broader multifaceted connection between migrants, their network and environment, and how familial relationships, intimacies and responsibilities have been facilitated (See Table 1).

For example, researchers such as Diminescu (2008), Hamel (2009) and Lim et al. (2016) emphasise how ICTs enable migrants to "circulate" while creating and maintaining near and distant connections. Sayad (2004) highlighted how ICTs are increasingly transforming the "twofold absence" of migrants to a dual and concurrent presence of 'here and there'. Be it simple "conversational" to substitute absence or "connected" to ensure a continuous virtual presence (Diminescu, 2008), this literature reflects migrants' continuous exploration and utilisation of ICT (Gifford & Wilding, 2013; Komito, 2011) including for familial relationships and intimacy.

Familial solidarities, shared commitments and familyhood have been manifested differently including by proxy—representational objects or persons compensating for the absent person, physically, or through imagined co-presence (Nedelcu and Wyss, 2016).These are extended and become more and more realistic amid advances in ICTs whereby emotions, care and support are being exchanged at a distance such as via live video calls. Some scholars have explored how ICTs are increasingly being used to "do" and connect to family, give care at a distance, and share resource, emotion, and responsibility.

For example, based on ethnographic research on Filipino women migrants in the UK, Madianou (2016a, 2016b) demonstrated how new ICTs—"polymedia"—enabled the "ambient co-presence" of families across multiple locations, and enhanced the "low-level emotional reassurance" for those with strong relationships. Similarly, qualitative research with Romanian migrants in Switzerland by Nedelcu and Wyss (2016) demonstrated how ICTs have shaped the "ritual", "omnipresent" and "reinforced" communications and their "ambivalent" effects such as feelings, emotions, and solidarity; and how ICTs enabled the "cosmopolitanization" of everyday life by enabling the capacity and feeling of "being and doing things together" at a distance (See Table 1 for more examples). Generally, ICTs offer tremendous benefits for migrants' multifaceted communications, and this has been demonstrated in a relatively large body of literature concentrated in the contexts and perspectives of the "Global North".

### 3.2. ICTs for facilitating the planning and execution of migratory journeys

Besides facilitating transnational and familial connections ICTs, primarily mobile phones, have also been praised for being essential "toolkits" particularly for "irregular" migrants during their often fragmented and dangerous migratory journeys. Some researchers have elucidated this interaction.

For instance, Gillespie et al. (2018), Dekker et al. (2018), Dekker & Engbersen (2014), Gough & Gough (2019) and Kang et al. (2018) denoted how ICTs enabled migrants to plan, navigate and document their journeys and to access services en route. Mobile phones have been used to calculate distances, learn about terrains, weather conditions and currencies, learn and translate languages, navigate places and services—refuelling points, money transfer and temporary shelters, illuminate dark footpath, connect to friends and family, humanitarian actors and smugglers. They served as "life savers" by providing a platform to signal and communicate dangers (Dekker et al., 2018; Mancini et al., 2019). They also provided digital alternatives to store and transport personal memories and crucial documents, maps, dictionaries, books, and cash which can make an immense difference for the migrants during difficult journeys. In sum, despite the associated challenges and risks (Section 4), as several researchers demonstrated (See Table 1) ICTs have become critical navigation, networking and "life-saver" tools for migrants during their often dangerous and fragmented migratory journeys.

### 3.3. ICTs for facilitating "integration"

This section overviews ICTs' role in migrant "integration". "Integration" is a hugely contested concept but is mostly construed as a dynamic, multidimensional, and a two-way adaptation in a "host" society (Ager & Strang, 2008; Fortunati et al., 2011a). Although limited to the "Global North" contexts and perspectives, there is significant emerging research on ICTs' role in migrant "integration"/inclusion which draws on the earlier "diaspora" literature. This can be classified in two overlapping categories.

The first category emphasises migrants' ICT-facilitated information practices pertaining to their information needs and enhancement of social capital for "integration". Although impact studies are generally lacking, this category explores migrants' ICT use—mainly mobile phones, internet, and social media—apropos different dimensions of integration like income and employment, language and culture, education and skills, and social capital.

However, views on the way ICTs influence "integration" are contradictory. One view foregrounds ICTs' positive role. For example, based on their analysis of resettled refugees' ICT practices in New Zealand, Andrade and Doolin (2018) identified eight ICT use patterns vis-à-vis well-being enhancement and "affective participation" including communication, online transaction and expressing cultural identity. Based on a qualitative analysis of Syrian refugees, AbuJarour and Krasnova (2017) indicated mobile phone use can enable the social inclusion of refugee migrants. Qualitative research by Alencar (2018) with 18 Syrian, Eritrean and Afghan refugees in The Netherlands highlighted the vitality of social media networking sites for language and cultural literacy and social capital development. Kaufmann (2018, p. 882) concluded "smartphones hold an untapped potential for integration processes" based on their analysis of Syrian refugees' ICTs practices using interviews, "WhatsApp" chats, and participant research. Aretxabala and Riezu (2013) concluded that ICTs increase social capital of migrants. "Mixed" method research by Felton (2015) with "non-English speaking migrants" in Australia revealed how mobile internet has been widely used for seeking accommodation and job-related location information.

The contrasting view highlights ICTs' negative role in "integration". For example, Suh and Hsieh (2019) in their interview-based research with Korean migrants in the USA revealed that while ICT-enabled information allowed the migrants to build and maintain "local" and "global" identity, the "intraethnic" communication slowed down their "integration". Komito & Bates's (2009) research revealed Polish migrants' use of social media and social networking technology reduces their "integration" in Ireland although it enables them to become "media rich and resilient". In their interviews and observation-based research with Turkish migrants in the United Kingdom, Aydemir (2018) posited that while ICT-enabled communication enhances bridging—connecting with members of different groups, it also "trivialises" the pre-emigration social ties, and increases "individualism and particularism". Wilding's (2009) analysis highlighted the unintended consequences of ICT-based projects aimed for the social inclusion of youth with refugee backgrounds in Australia—misuse of personal data and protraction of obligations and responsibility.

The second category relates to the development, implementation, and outcomes of ICT-based projects apropos "integration"/inclusion. For instance, Codagnone & Kluzer (2011) synthesised five studies that examined ICT-based initiatives relating to immigrants and ethnic minorities (IEM) in 27 European countries and 11 case studies from France, Germany, Spain and the UK using survey, case analysis and workshops. The findings highlight a user-driven and bottom-up ICT uptake by IEM and the importance of third sector entities and IEM associations to enable ICT use for inclusion. The study also highlights the hindrances of digital inequality for socio-economic inclusion of IEM as a result of factors such as socio-economic status, gender, language, infrastructure and inadequate digital content. Brown and Grinter (2016) reported a "human-in- the-loop" interpretation messaging platform that mediates the communication between two groups with different languages. Almohamed et al. (2018) illustrated a promising participatory approach to designing ICT solutions for refugee "integration" which involved NGOs, volunteers, refugees and ICT designers to inform the design concept. Muñoz et al. (2018) explored the potentials of free digital learning (FDL) tools and initiatives for migrants' language learning, civic integration, employment, and higher education. Overall, despite the lack of impact research and the varying views on ICTs' actual role in migrants' "integration", there is significant emerging research in this domain (See also Table 1).

### 3.4. ICTs for enabling livelihoods

Livelihoods is another area affected by ICTs. The livelihoods of migrants and their families are highly enmeshed with aspects of migration/development including remittances and investments (Duffield, 2006; Sørensen et al., 2002) which are increasingly entangled with ICTs (Hackl, 2018; Steel et al., 2017). However, very little is known about the contemporary interplay between ICTs, livelihoods and migration/(im)mobility/development particularly in the "Global South" context (Scoones, 2009; Steel et al., 2017). Existing studies tended to overlook ICTs' role; concentrate on rural households with a narrowed view of "naïve localism" (Scoones, 2009); and view livelihood enhancement only as income increment (Czaika and De Haas, 2014). The disregard of ICTs' role has been contradicting people's increasing use of these technologies in their daily livelihood-related activities. In fact, as Heeks (2006) underlined, this issue is prevalent in the wider ICTs and development—"there has been a bias to action, not a bias to knowledge. We are changing the world without interpreting or understanding it" (p. 1).

However, some broadly related limited work in ICTs and migration-related livelihoods—income, employment, and skills; remittances; and livelihoods amid environmentally-related displacement are worth mentioning. Regarding the former, for example, Steel et al. (2017) in their case studies in Cameroon, Rwanda and Sudan found that digital connections and the flow of people, goods, services and information are improving livelihoods. However, some are affected by digital inequality and the politics of mobility (Cresswell, 2010)—unable to move or access new technologies due to social inequalities—"trapped populations"(Boas, 2017). Those without access to mobile phones become more and more dependent on traditional intermediaries for job opportunities and market information. Some grey literature also featured digital labour case studies relating to refugees and IDPs like "Samasource"[1] (Hatayama, 2018; Samasource, 2019). Hackl (2018, p. 158) states "(…) 'digital livelihoods' are already becoming a new development mantra […] refugees should tap into digital economies by becoming data cleaners and crowd workers, thereby building a brighter future from cash for food to tech for food". Nevertheless, such optimism appear to be misplaced as the premise represents only limited groups—young technology savvy and relatively wealthy Syrian refugees (Borkert et al., 2018; Kaufmann, 2018) as denoted in Mansour's (2018) qualitative study which shows that two-thirds of the Syrian refugees who crossed to Egypt have high digital literacy skills.

Hatayama (2018) reported how internet, mobile phones, applications, and platforms enabled some refugees and IDPs to enhance income and financial resources—including by delivering online language training, crowdfunding, peer-to-peer lending, e-commerce, and entrepreneurship; and also, to access and share information and to develop skills—including information about local employment opportunities and regulations, and to access content and teachers from higher education institutions remotely. There is also a growing trend of delivering computer coding and other IT skills to refugees and IDPs in refugee camps (Hatayama, 2018). Fintech and biometric solutions such as blockchains and biometric identity cards are also portrayed as assisting refugees and IDPs to access financial and essential services. However, it unclear whether and how the notorious privacy and ethical issues related to the use of such technologies in the context of vulnerability (Madianou, 2019; Maitland, 2018; Thomas, 2005) have been addressed. Hatayama's (2018) report appears to represent only limited cases since such applications are largely absent in many contexts including vast refugee camps in the Horn of Africa. Mobile money systems such as M-PESA in Kenya have been reported to facilitate payments and financial transactions to the poor including "asylum seekers" and refugees in camps who mostly are deprived of traditional financial services (Miao et al., 2018; Suri & Jack, 2016).

Regarding remittances and livelihoods amid environmentally-related displacement, research on remittances tended to focus on households in migrant sending countries, and ICTs' role has been largely ignored. Although specific research on ICTs' role in livelihoods during environmentally-related displacement is generally lacking (Boas, 2017), some researchers have highlighted the importance of ICTs in such situations. For example, Lu et al.'s (2016) analysis of the mobility of people affected by Cyclone Mahasen in Bangladesh using mobile phone data indicates the usefulness of mobile phones to understand fast-changing human mobility which can offer crucial information on people's livelihood and safety in the time of disaster. Hiremath and Misra (2006) presented how ICT-enabled migration information intervention in Gujarat, India addressed critical local livelihood problems. They stated a simple migrant information centre enabled by wireless in local loop (WLL) telephony enormously supported the social and economic security of the vulnerable migrants by reducing migration costs, enhancing communication and networking, improving employment opportunities, protecting labour rights, improving perception towards migrants, and enhancing emotional, food and financial security. If the project delivered as claimed, the success appears to rest on its identification of local needs and participatory appraisal.

---

[1] "Samasource", a digital microwork service by a San Francisco based company, is reported to secure large and labour-intensive data and artificial intelligence (AI) related digital work contracts such as web search and evaluation, image tagging and transcription from corporate companies in developed countries such as Google, CISCO and Yahoo. The company then decomposes, encapsulates and distributes the micro digital work using online platforms to people in poverty including refugees and IDPs in camps in Kenya, Uganda, India, Haiti and Costa Rica (Borokhovich et al., 2015; Gino & Staats, 2012a; Hatayama, 2018). It has been stated that the business model connects the cheap labour in developing countries to the high labour demand in countries with advanced economies and to use microwork to benefit marginalised people to earn a living wage instead of receiving aid (Borokhovich et al., 2015).

In summary, apart from the broadly related studies cited earlier (See also Table 1), research targeting ICTs' role in the contemporary migration-related livelihoods is rare. This is despite the strong entanglement of migration, livelihoods, and ICTs.

| Research area in ICTs and migration | | Example relevant literature |
|---|---|---|
| ICTs in "diasporic" and transnational connection | Connection between migrants, and their broader network and environment | Andrade & Doolin, 2018; Awad & Tossell, 2019; Boas, 2020; Collin et al., 2015; Dekker et al., 2016; Diminescu, 2008; Fortunati et al., 2011b; Georgiou, 2006; Gifford & Wilding, 2013; Hall, 1990; Hamel, 2009; Komito & Bates, 2011; Larsen et al., 2006; Levitt, 1998; Levitt & Jaworsky, 2007; Lim et al., 2016; Marlowe, 2019; Oiarzabal, 2012; Oiarzabal & Reips, 2012; Pearce et al., 2013; Thompson, 2009; Vincent, 2014; Wilding, 2006; Wyche & Chetty, 2013 |
| | Familial relationships, intimacies, and responsibilities | Acedera & Yeoh, 2019; Bacigalupe & Lambe, 2011; Baldassar, 2008, 2016a, 2016b; Baldassar et al., 2006; D. Brown & Grinter, 2012; R. H. Brown, 2016; Cabalquinto, 2019; Chib et al., 2014; Cuban, 2014; Elliott & Urry, 2010; Francisco-Menchavez, 2018; Francisco, 2015; Horst, 2006; Kutscher & Kress, 2018; Madianou, 2012a, 2012b, 2016a, 2016b; Madianou & Miller, 2011, 2013; Merla & Baldassar, 2016; Nedelcu, 2012, 2017; Nedelcu & Wyss, 2016; Peng & Wong, 2013; Robertson et al., 2016; Vertovec, 2004 |
| ICTs in migrants' "integration" | | AbuJarour et al., 2019; AbuJarour & Krasnova, 2017; Alam & Imran, 2015; Alencar, 2018; Alencar & Tsagkroni, 2019b; Almohamed et al., 2018; Andrade & Doolin, 2018; Aretxabala & Riezu, 2013; Aydemir, 2018; D. Brown & Grinter, 2016; Codagnone & Kluzer, 2011; Collin & Karsenti, 2012; Felton, 2015; Garrido et al., 2010; Kaufmann, 2018; Komito & Bates, 2009, 2011; Muñoz et al., 2018; Siddiquee & Kagan, 2006; Suh & Hsieh, 2019; Wilding, 2009 |
| ICTs in behaviours and processes of migration | Migration aspirations and decisions | Alampay, 2018; Andrade & Doolin, 2018; Boas, 2017; Borkert et al., 2018; Cooke & Shuttleworth, 2018; Czaika & De Haas, 2014; Czaika & Parsons, 2017; Dekker et al., 2018; Dekker & Engbersen, 2014; Elbadawy, 2011; Frouws et al., 2016; Haug, 2008; Kotyrlo, 2019; Latonero, 2012; Leurs & Smets, 2018; Madianou, 2012a; Mancini et al., 2019; Molony, 2012; Morrison & Clark, 2016; Onitsuka & Hidayat, 2019; Poot, 1996; Smets, 2018; Spaan & van Naerssen, 2018; Thulin & Vilhelmson, 2014; Vilhelmson & Thulin, 2013 |
| | Migratory journeys | |
| | | AbuJarour & Hanna, 2017; Alam & Imran, 2015; Alencar, 2018; Alencar et al., 2019; Alencar & Tsagkroni, 2019; Almohamed et al., 2018; Andrade & Doolin, 2018; Aretxabala & Riezu, 2013; Aydemir, 2018; Borkert et al., 2018; D. Brown & Grinter, 2016; Chib & Aricat, 2017; Codagnone & Kluzer, 2011; Collyer, 2010; Dekker et al., 2018; Dekker & Engbersen, 2014; Felton, 2015; Frouws et al., 2016; Garrido et al., 2010; Gillespie et al., 2016, 2018; Gough & Gough, 2019; Hannides et al., 2016; Hashemi et al., 2017; Holmes & Janson, 2008; Kang et al., 2017; Kaufmann, 2018; Komito & Bates, 2009; Maitland, 2018; Mancini et al., 2019; Muñoz et al., 2018; Newell et al., 2016; Siddiquee & Kagan, 2006; Suh & Hsieh, 2019; Vernon et al., 2016; Wei & Gao, 2017; Wilding, 2009 |
| ICTs and migration-related livelihoods | | Borkert et al., 2018; Borokhovich et al., 2015; Gino & Staats, 2012b, 2012a; Hackl, 2018; Hiremath & Misra, 2006; Kaufmann, 2018; Latonero & Kift, 2018; Lu et al., 2016; Miao et al., 2018; Patil, 2019; Steel, 2017; Steel et al., 2017; Suri & Jack, 2016; Yafi et al., 2018 |
| Migrants and digital risks | Migrants and digital visibility, surveillance, and border control | Alencar et al., 2019; Beduschi, 2017; Chouliaraki & Musarò, 2017; Frouws et al., 2016; Gillespie et al., 2018; Hugman et al., 2011; Jumbert, 2018; Jumbert et al., 2018; Kingston, 2014; Latonero & Kift, 2018; Madianou, 2019; Maitland, 2018; Mancini et al., 2019; Newell et al., 2016; Thomas, 2005; Vannini et al., 2019; Vukov & Sheller, 2013; Wall et al., 2017 |
| | ICTs, human trafficking, and extortion | Borkert et al., 2018; Dekker et al., 2018; Beşer & Elfeitori, 2018; Harney, 2013; Latonero, 2011; Latonero & Kift, 2018; Maitland, 2018; Newell et al., 2016; Mirjam Van Reisen & Mawere, 2017; Van Esseveld, 2019; Vukov & Sheller, 2013; Wall et al., 2017 |
| ICT-enabled communication, burdens, and pressures on migrants | | Akuei, 2005; Awad & Tossell, 2019; Horst, 2006; Madianou, 2016b; Nedelcu & Wyss, 2016 |
| Migrants and "digital divide" | | Alam & Imran, 2015; Benítez, 2006, 2010; Caidi et al., 2010; Chatman, 1996; Clark & Sywyj, 2012; Frouws et al., 2016; Fuchs, 2009; Goodall et al., 2010; Harney, 2013; Katz et al., 2017; Leung, 2018; Platt et al., 2016; Robertson et al., 2016; Salman & Rahim, 2012; Vertovec, 2009; Wilding, 2009; Yu et al., 2018 |
| Digital migration study | | Dekker & Engbersen, 2014; Kotyrlo, 2019; Leurs & Prabhakar, 2018; Leurs & Smets, 2018; Lu et al., 2016; Messias et al., 2016; Sánchez-Querubín & Rogers, 2018; Schrooten, 2012; Zagheni et al., 2014 |

**Table 1: Key research areas and relevant literature in ICTs and migration**

# 4.  THE TYRANNY OF ICTS FOR MIGRANTS

As noted earlier, ICTs are often viewed as liberating and utilised at migrants' discretion. As such, most existing literature emphasises their benefits. However, besides their benefits, these technologies also pose critical risks, burdens, pressures, and inequality. Although specific research on the tyrannical aspects of ICTs is extremely rare, this section sheds light on this important area by drawing on reflections from the wider related literature. Three key areas are particularly emphasised and discussed in turn: digital risks; ICTs-related burdens and pressure; and digital inequality.

## 4.1.  Migrants and digital risks

Apart from risks that they share with all other ICT users such as computer-induced health risks, digital risks associated with migrants can be viewed in two major lenses. They are digital visibility, surveillance, and border control on the one hand, and human trafficking and extortion on the other as elucidated below.

### 4.1.1. Migrants and digital visibility, surveillance, and border control

There are increasing risks associated with ICT-enabled visibility and surveillance of migrants primarily when they flee home and travel within and across borders in pursuit of safer places (Mancini et al., 2019; Wall et al., 2017). These include surveillance by source and destination nation states, smugglers, traffickers and criminal gangs (Latonero & Kift, 2018; Vukov & Sheller, 2013), risks linked to the use of digital data such as digital biometric identifications which may expose migrants' identities (Maitland, 2018), and data generated as a result of migrants' use of ICTs such as mobile phone and social media usage data—often quoted as "big-data".

The post 9/11 widespread use of surveillance technologies, "technologies of control", at borders and beyond has been exacerbated amidst the ICT-assisted "irregular" migration notably since the 2015/16 "refugee crisis". Some researchers have highlighted the increasing sophistications of these technologies and how their use is increasingly deterring the rights of protection and privacy of vulnerable migrants (Jumbert, 2018; Madianou, 2019; Vannini et al., 2019). Digital borders have extended beyond physical borders whereby large amounts of migrants' data—often racially profiled—are collected and stored in large, distributed databases and shared among several actors. In addition, many European immigration and border authorities have sought and gained legal impunity to search (and even to hack) migrants' phones (Frouws et al., 2016; Jumbert et al., 2018). While such activity may inform some elements of the intended purposes—such as verifying and identifying migrants from other security threats, overreliance on such technologies may also lead to erroneous and risky outcomes—for example, migrants may self-censor; mobile phone use may have been shared among many users; and migrants' online identity may be hacked and faked by others.

Some studies highlighted different forms of these risks. For example, Beduschi (2017) highlighted migrants' risk of being tracked and discovered by traffickers, smugglers, border guards and common criminals via global positioning system (GPS) applications. Chouliaraki and Musarò (2017) observed how authorities in Greece and Italy used ICTs to intercept migrants before they reach borders. In their study of Syrian and Iraqi refugees' journey to France, Gillespie et al.(2018) highlighted the tension between the importance of mobile phones as "lifelines, as important as water and food" and mobile phones being new sources of exploitation, abuse and surveillance for migrants as migration has been criminalised by European policy  and financialised by smugglers. Mobile phones have sometimes been a source of torture and even death such as when migrants' possession of sensitive information is discovered by perpetrators.

Other studies stressed concerns regarding the growing use of biometric identification systems in relation to vulnerable migrants by various parties including governments, and humanitarian and private organisations (Kingston, 2014). Biometric identifications can easily expose individuals as they involve physiological and behavioural patterns of a person such as fingerprint, DNA, iris, face, voice, ear shape, body odour, hand-written signature, keystroke and gait (Kingston, 2014; Thomas, 2005). Jumbert et al. (2018, p. 4) stressed that:

> *Of particular concern are the implications of a hybrid form of governance, where private companies, state authorities, NGOs and international organizations fail to understand the surveillance capabilities of digital devices or fail to set high standards of digital safeguards.*

Furthermore, migrants' increased use of ICTs has produced abundant digital data that can characterise migrants' online behaviour as well as their identities. While ethical, cautious and responsible use of such data could be beneficial in many ways such as "for the protection of migrants' human rights by enhancing both decision-making and measures to prevent unnecessary deaths at sea, ill-treatment and human trafficking of migrants" (Beduschi, 2017, p. 981), inappropriate use of such data could also risk migrants' life and privacy. However, research examining the extent of risk posed or exhibited on migrants and their networks as a result of inappropriate use of migrants' digital data is lacking.

Nevertheless, it should be noted that migrants are not always passive victims of surveillance as they sometimes appropriate their use of ICTs. For example, Gillespie et al. (2018) documented how Syrian refugees used ICTs creatively to avoid surveillance such as by protecting their digital identities, using encrypted mobile applications, closed groups, pseudo-names and avatars in social media to share sensitive information. In short, the increasing use of "technologies of control" and migrants' digital traces by various actors poses a serious risk for vulnerable migrants despite this important issue is raised only in passing in most of the existing literature (See Table 1).

**4.1.2. ICTs, human trafficking, and extortion**

Digital visibility and "technologies of control" are not the only digital risks for migrants. Some researchers reflected on how ICTs primarily mobile phones and social media have been used by smugglers, traffickers and gangs to misinform, enslave, extort and torture migrants for ransom (Latonero, 2011; Van Esseveld, 2019) as well as how migrants may consume misleading and fake information from social media which sometimes may lead to deadly consequences (Borkert et al., 2018). Van Esseveld (2019) and Reisen and Mawere (2017) shed light on how mobile phones are being used by traffickers and armed groups in transit countries such as Libya to abuse and extort migrants. As highlighted in "Tortured on Camera: The Use of ICTs in Trafficking for Ransom" (Van Esseveld, 2019), traffickers have used live calls while migrants are being tortured and screaming, recorded images and videos of torture to collect ransom from distant families and friends.

Beşer & Elfeitori (2018) underline how European partnership with Libya to prevent migrants from coming to Europe has encouraged violations including arbitrary detention, enslavement, malnutrition, the lack of hygiene, torture, and sexual abuse in Libya. Newell et al. (2016) in their study of information practices of undocumented migrants at the USA-Mexico border highlighted how possessing contacts of friends and family subjected migrants to risks of extortion and abuse by drug and human traffickers, thieves, and corrupt officers; and how migrants mistrusted and rejected ICT-based assistance including from humanitarian actors in fear of surveillance. Further, migrants' reliance on locals for basic services such as mobile cellular subscription due to lack of essential documents and legal rights subjected them to fraud, blackmailing and exploitation (Dekker et al., 2018; Newell et al., 2016; Vannini et al., 2019). In sum, although specific research is rare, emerging evidence shows the increasing use of ICTs for human trafficking and migrant extortion (See Table 1).

## 4.2. ICT-enabled communication, burdens, and pressures on migrants

Migrants' increased ICT-enabled communication and connectedness come with costs of burden and pressure like social surveillance, obligation and commitments (Awad & Tossell, 2019). For instance, Horst's (2006) field study in Jamaica highlighted how increased access and use of mobile phones leads to unforeseen burdens and social obligations whereby distant migrants are "burdened by love and the compunction to give" and increased gendered surveillance and involvement in each other's everyday life. Akuei's (2005) study on Sudanese resettled refugees in the USA shows how remittances as social obligations pose unforeseen pressure on resettled refugees. "Knowing what their kin are facing in Cairo and other asylum locales, resettled refugees expressed feeling a deep responsibility to honour these obligations. […] with their mounting remittances and meagre resources to improve their own lives, it is the resettled refugees who now feel the poorer" (p. 13). Nedelcu and Wyss (2016) denoted how ICT-facilitated co-presence not only supplements the wellbeing and quality of family relationships, but also its "ambivalent" effect adds extra pressure and commitments primarily on distant migrants to respond to calls at any instant. ICTs also enabled increased cultural surveillance whereby those with weak relationships experienced increased conflict (Horst, 2006; Madianou, 2016b). Therefore, alongside the benefits, ICT-enabled communication and connectedness bring costs of burden and pressure on migrants.

## 4.3. Migrants and digital inequality

Digital inequality is another key challenge for vulnerable migrants. Digital inequality can be understood as differences in accessing, using, and benefiting from ICTs among various groups due to factors such as power and economic status, rights, age, sex, and geographical location. Earlier studies focused on a "digital-divide" which emphasised variations in economic participation and physical access to technology such as personal computers (Duncombe & Heeks, 2002; Warschauer, 2002). While still mainly concentrated on "sociodemographic and socioeconomic determinants" (Scheerder et al., 2017, p. 1607), the contemporary notion of "digital-divide" is beyond mere access to technological equipment (DiMaggio & Hargittai, 2001; Warschauer, 2004). Here, 'digital inequality' instead of "digital-divide" is used to reflect the multidimensional inequalities driving and driven by the uneven distributions of not only the technologies but also the motivation, resources, structures and capabilities to use them (Fuchs, 2009). Besides the technology, other dimensions—for example, behavioural (motivation and skills) and usage-related (e.g., bandwidth and frequency of use)—are critical (van Deursen & van Dijk, 2019). In fact, digital inequality may exist where physical access is universal (van Deursen & van Dijk, 2019). Belloni's (2020) ethnographic study with Eritrean refugees in Italy and Eritrea showed that despite having access to technology refugees have been avoiding communication with home in response to the overwhelming expectations and responsibility until they reach their aspired upward destinations and settle. Thus, as Norris (2001, p.32) emphasised, digital inequality should be analysed in all three levels – micro (resources and motivations at individual level), meso (political and institutional contexts) and macro (economic and technological contexts).

While digital inequality is not limited to migrants, it needs a contextual analysis given migrants' heightened information needs. The purpose and use of ICTs by migrants such as asylum-related migrants is significantly different from the rest of the society; and migrants have specific information needs often characterised as information poverty (Chatman, 1996) and challenged by structural and social barriers (Caidi et al., 2010; Maitland, 2018). Due to their particular situations, asylum-related migrants are often very information precarious which could lead them to misinformation and vulnerability (Harney, 2013). Leung's (2018) study in Australia, for example, shows that while refugees and "asylum seekers" use the internet mainly to access information relevant to their challenging circumstances and to connect to their families and fellow migrants, "netizens" (mainstream users in the general society) use internet mainly for online presence. The author also revealed the legitimacy to use ICTs is dually questioned (by origin and destination countries) for refugees/asylum seekers (like most social, legal and economic resources) whilst it is authorised for "netizens"—as long as the authority of the nation-state is not challenged. Merisalo & Jauhiainen's (2019) research on the ICT use of migrants from 37 countries in Africa, Asia, and Middle East in their journeys from origin country to transits and destinations demonstrated how the circumstances of asylum-related migrants forced them to use ICTs although they had not had prior habit, interest, and skills.

While specific and impact research on migrants' digital inequality is rare, some studies have considered variations in access and use of ICTs by migrants and families with migration backgrounds vis-à-vis factors such as age, gender, income and education either with respect to origin country (Benítez, 2006; Merisalo & Jauhiainen, 2019) or recipient society (Leung, 2018; Yu et al., 2018) (See also Table 1). Leung (2018) highlighted how English-only internet content and affordability exacerbated the "digital-divide" among refugee background users in Australia. Katz et al. (2017) highlighted how low digital access and use affected the learning and development of "immigrant" children and "low-income" families in the USA. Benítez (2006) highlighted the importance of skills, education, family, and cultural capital for the successful use of ICTs; and how ICTs are more vital for those migrants constrained to travel by legal, economic, and social factors. Some studies explicated the role of social relations in digital inequality. For instance, Platt et al. (2016) highlighted how Indonesian domestic workers' ICT-enabled communication in Singapore is contingent on the "always ongoing" negotiation with their employers as their communication can be deterred by way of surveillance and obscuring "WiFi" passwords. In sum, multilevel and multifaceted digital inequalities are increasingly affecting vulnerable migrants. However, although some studies conveyed variations in migrants' access and use of ICTs explicitly or implicitly, they did not adequately problematise the subject and its consequences as important areas of research.

## 5. GAPS AND OPPORTUNITIES

Sections 3 and 4 examined the status quo of literature relating to ICTs and migration apropos the liberalities and tyrannies of ICTs for vulnerable migrants. This section highlights eight insights, gaps and opportunities relating to both the existing and missing objects of research in ICTs and migration/development.

First, existing research on ICTs and migrants is almost entirely limited to migrants' ICT use 'en route' or within the "Global North" and perspectives from these regions. Thus, it needs decolonising and reimagining vis-à-vis the "Global South" as not least the socio-economic, cultural, technological, and political contexts of the "Global South" are significantly different from the "Global North", and over 80 per cent of internal and international migration takes place within regions in the "Global South".

Second, despite some existing research on migration- and mobility-related livelihoods—seasonal and local migration/mobility, remittances, refugees' self-reliance, etc., there are several gaps. Notably, most existing studies are place-deterministic—failing to consider the multilocality of livelihoods particularly in transnational contexts and overlooking livelihood's multidimensionality—they tend to focus merely on income. They also ignore ICTs' role despite the increasing entwinement of people's movements and ICT use apropos diversifying and securing livelihoods often across multiple locations. Despite some challenges[2], ICTs can extend corporeal movements to imaginative, virtual, and communicative ones whereby enabling the physically immobile—those bound by socio-economic constraints, family responsibility, age, cultural norms, etc., to access livelihood opportunities at a distance. However, research that recognises ICTs' role vis-à-vis the diversity and multilocality of contemporary livelihoods and the peculiar circumstances of migrants particularly in the "Global South" context are exceptionally scarce. This can be seen as important opportunity for future research.

Third, while ICTs have become migrants' critical navigation, networking and 'life-saver' toolkits; and enabled migration management for several actors, they also pose risks primarily for vulnerable migrants. Misuse of migrants' online and usage data and biometric identification systems poses serious risks for these migrants. Research investigating the extent of such risks in different contexts particularly from the perspective of migrants and exploring a balanced way of developing and using these technologies by putting migrants' lives and their participation at the centre is needed.

Fourth, literature indicates that ICTs are increasingly enabling the realisation of a virtual relationship and intimacy nearly typical of in-person interactions particularly for those constrained by distance. Nevertheless, they also pose pressure and burdens particularly on distant migrants laden with multiple socio-economic obligations and commitments. However, existing studies tend to view ICTs only optimistically; and critical research cogitating both the 'bright' and 'dark' consequences of ICTs particularly on vulnerable migrants is generally lacking. This is one important gap to be addressed by future research.

Fifth, there is sufficient evidence confirming the increasing influence of ICTs on migration behaviours, patterns and processes including (potential) migrants' aspirations and decisions (See Table 1). However, existing evidence is inadequate to answer very important questions including 'how increase in access and use of ICTs relate to inward/outward migration?' necessitating contextual and focused research particularly vis-à-vis the growing informal, fragmented and often dangerous migration apropos both the source and destination countries at micro, meso and macro level. Future research in this area can benefit from drawing on existing migration theories such as new economic migration theory[3] to understand the role of ICTs in migration. However, such theories need to consider the role of power and gender in collective decision making.

Sixth, while several researchers have attempted to explore the facilitating role of ICTs in migrants' "integration", existing studies not least focused only on ICTs' positive role often only in the perspective of the isolated contexts in the "Global North". Critical research on ICTs' role in and impact on migrants' multidimensional inclusion especially in relation to the digital and the wider multidimensional inequality is needed.

Seventh, migrants are increasingly affected by digital inequality which has broader implications to other types of multilevel and multidimensional inequality. Although some researchers have conveyed variations in migrants' ICT access and use explicitly or implicitly, they have failed to adequately problematise the subject and its consequences as important areas of research signalling future attention.

---

[2] Those who lack the means and capability to move either physically or virtually are unable to leverage the potentials of mobility and migration for their livelihood advantages. These immobility traps and their consequences can be caused by deficiencies associated with both physical and virtual mobility. In addition to constraints of physical mobility such as financial resources, those who are in the remote side of the digital-inequality – those constrained by factors associated with physical access, motivation and capability to use ICTs are often excluded from the benefits of virtual mobility. However, less mobility does not always infer more immobility and sometimes mobility also promotes immobility. For example, immobility can be reinforced by the mobility of others such as households of migrants due to domestic and caretaking responsibility (Kothari, 2003; Steel et al., 2011). Other times immobility may be strengthened by social obligation of migrants such as responsibility for remittances (Akuei, 2005; Bailey et al., 2002; Hunter, 2015).

[3] New economic theory of migration argues that decision to migrate is not a result of a decisions of isolated individual actors to maximise wage and income; but it might be a collective decision not only to maximise income and wage but also to minimise or disperse risks (de Haas, 2010; Stark, 1985; Taylor, 1999). This theory also holds that migration can lead to a relative deprivation where income inequality in a community as a result of migration and remittance might lead to further migration (J. E. Taylor, 1999); and increase in local employment, production and wages does not necessarily stop migration. The older neoclassical economic theory of migration assumes geographical labour market and wage differentials, and individuals' rational choice to maximise utility determine migration motives and decisions both at micro and macro level (Harris & Todaro, 1970; King, 2012; Massey et al., 1993; Sjaastad, 1962; Todaro, 1969).

Finally, ICTs are increasingly influencing migration research. They are reshaping data collection and analysis methodology with important implications for migration theorisation (See 'Appendix II' and Table 1). Rigorous research on the ethical use of digital methods for migration studies is another important opportunity for future research.

## 6.    CONCLUSION

In a bid to assess the current status quo and illuminate future directions in ICTs and migrants research, this paper has attempted to outline contemporary knowledge and pinpoint the key gaps and future opportunities particularly vis-à-vis ICTs' liberating and tyrannical role especially for vulnerable migrants like the displaced and trafficked. This was done by drawing on 157 published materials relating to ICTs and migration/development. Against the dominant view on ICTs' empowering role, the analysis elucidates the "double-edged sword" consequences of digital technologies by expounding how these technologies not only offer opportunities but also pose risks, burdens, pressures, and inequality predominantly on vulnerable migrants. It establishes how ICTs, particularly mobile phones and the internet, are increasingly reconfiguring all aspects of migration including migrants' networking and connections, behaviours and processes of migration, migrants' integration and social capital, and migration-related livelihoods. In doing so, however, these technologies not only empower these migrants such as by facilitating dangerous migratory journeys and social connections but also subject them to risks including ICT-facilitated trafficking, torture, extortion, and surveillance as well as burdens and pressure including protracted responsibility and commitments and multifaceted socio-economic inequality created or amplified by ICTs. Overall, the analysis provides 'food for thought' particularly on the need to thoroughly consider both the contextual 'bright' and 'dark' role and consequences of ICTs for a more realistic and nuanced ICTs and migration research and practice.

# APPENDIX I: LITERATURE SELECTION AND THEMATIZATION APPROACH

1. **Data source and collection**
1.1. **Source of data**

The review process began with searching databases, academic journals, and conferences with mixed combinations of keywords (See the 'Keyword and searching strategy' section) from fields of new media and communication, information systems, information and communication technologies (digital technology), information and communication technologies for development (ICT4D), migration, development studies, law, geography, economics, sociology, anthropology and history. The literature search and identification process primarily focused on a peer-reviewed published academic research. However, other relevant publications such as practitioner reports, policy documents, working papers from authoritative sources and thesis have been also included in the review process.

Search engines, databases, and journals including Google Scholar, ABI/Inform collection via ProQuest, Scopus, Web of Science, ACM Digital Library, ScienceDirect, JSTOR, and relevant sources in the list of 79 ICT4D-related publication resources collated by (ICTworks, 2012) were used to search and retrieve electronic materials. In addition, peer-reviewed published works which emerged from the analyses of forward and backward cross-referencing have been included in the review. The University of London's multi-institution online catalogue, the Senate House library and the Royal Holloway University of London's library collections have been used as primary sources for the location and retrieval of relevant paper-printed publications such as books for which digital copies were not available or inaccessible. In some cases when very relevant full text publications were not accessible, authors are directly approached for a copy via academic networking platforms such as ResearchGate and Academia. Due to the multidimensional and interconnected nature of the constructs of digital technology, migration, inequality/development, this review is by no means considered to be exhaustive.

The keyword search (See the next section) returned tens of thousands of search results which mostly originated from migration, diaspora, transnationalism, mobility, sociology, ICTs for Development (ICT4D), development studies and media and communication studies. The publications were sourced from the following journals and publication sources: *Information Technology and people, Global Networks, Social Media and Society, Information Communication and Society, Popular Communication, Journal of Computer-Mediated Communication, Ethnic and Migration Studies, Information Technology and Society, Media, Culture and Society, MIS Quarterly, Critical Sociology, Gender and Society, Migration and Development, Family Issues, New Media and Society, Mobile Communication, Information Technologies and International Development, Information Systems Journal*, conferences and publications from the *Association of Computing Machinery*.

### 1.2. Keywords and searching strategy

To achieve the searching strategy, combinations of terms representing concepts of digital technology, migration and inequality/development were searched using search engine friendly Boolean operators. The initial attempt was to retrieve publications named by the combination of at least three concepts in the nexus of digital technology, migration, inequality, and development such as 'ICT and migration and inequality or development'. However, this was not sufficient to capture all related literature. Consequently, a range of combination of keywords representing at least two concepts simultaneously in the nexus, for example, 'ICT and migration', 'digital technology and inequality', 'ICT and inequality', 'migration and development', 'migration and inequality', 'digital inequality' were added to the search strategy. The list of keywords used in the search to represent the relevant concepts are summarised in 'Table a'.

| Concept aimed | Keyword used |
|---|---|
| **ICTs** | (digital) technology/technologies, ICT, social media, mobile phone, internet, (new) media, (new) technology/technologies, information and communication technology/technologies, ICT4D, information system |
| **Migration, migrants and mobility** | migration, migrations, migrant, migrants, mobility, refugee, global south migration, south-south migration, displacement, transnational, diaspora, forced migrants/migration, mobility |
| **Inequality** | inequality, marginality, exclusion, inclusion, discrimination, digital divide, digital inequality, e-inequality, e-exclusion, e-inclusion, digital inclusion |
| **Development** | international development, human development, global development, development |

**Table a: Keyword searched in the literature**

**1.3.     Selection of publications**

The selection process of the identified literature underwent three stages. In the first stage, 1314 publications were identified and their citation information and digital full text whenever available and accessible was downloaded and recorded to Mendeley reference manager. The choice of Mendeley over other competing reference manager software such as Zotero, RefWorks, CiteULike and EndNote for this research was mainly due to a personal preference as well as its features such as PDF format friendliness, generous free cloud storage (currently 2GB), desktop and web versions of the application, good synching feature allowing work from multiple locations and devices (Parabhoi et al., 2017). At this stage, publications were considered based on search engine indexing information such as title, source link, reasonable number of citations in accordance with publication year. Publications that were older than 30 years but had no evidence of any citation in the database that listed the publication or Google Scholar were not considered. In the second stage, 870 publications were selected based on a review of the abstract or executive summary whenever available or a quick skimming of headings and content pages for relevance. In the third stage, 157 publications were selected based on a further review of the papers for perceived quality by reviewing the combination of the abstract or executive summary, full or part of the introduction page based on the length of the publication, chapter and topic headings, full or part of the methodology and conclusion sections.

**2.     Scope and in/exclusion criteria**
**2.1.     Publication year**

Due to the rapid changing nature of ICTs and their resulting implications for migration, inequality, and development, only materials published after 1990 were considered in this review.

**2.2.     Publication language**

For this review, only publications in English were considered. This is due to limitation to translate publications in all other languages to English within the limited time and resource available and a deliberate reluctance to use automatic translators due to quality concerns.

**2.3.     Interdisciplinary nexus**

The choice of publications for inclusion based on their coverage of any of the dimensions in the digital technology, migration and inequality/development nexus was shaped by two major constraints. On one hand, the concepts of digital technology, migration and inequality/development are very broad and complex by nature and considering publications from all of these disciplines per se for review was unfeasible and challenging for analysis. On the other hand, studies that span over the intersection of all these disciplines are limited in the literature. Consequently, literature that intersects across at least two dimensions in ICTs, migration and inequality/development nexus were considered in the review to reconcile the situation. These intersections were ICT-migration-inequality/development, ICT-migration, ICT-inequality/development, migration-inequality/development.

## 2.4. Type of migrant/migration

For the purpose of this review, although there may be overlap in rare cases, only literature that engages with international migration (migration across national borders) rather than internal migration or displacement within a country has been considered. In considering the literature, no distinction has been made between various aspects of migration such as pre-migration, journey, settlement and return and types of migrants such as labour or economic migrants, forced migrants such as refugees and "asylum seekers" and other migrants such as student migrants. In addition, no type or aspect of inequality and development was predefined for the purpose of the literature search.

## 2.5. Geographical coverage

As the research informed by this review primarily aims to investigate the link between ICTs, migration, and inequality/development within countries of developing economies, initially, the review was piloted merely on the context of the developing economies or the "Global South". However, unfortunately, this generated very few materials. Hence, since the primary purpose of this review is exploration of the theories, methods, and empirical evidence in the existing literature to understand the status quo rather than drawing statistical generalisation, publications, and studies of various nature from a range of geographies and contexts have been considered.

## 2.6. Publication quality

For the in/exclusion of the literature for analysis, most aspects of the 'concept hierarchy of research quality' framework proposed by Mårtensson et al. (2016, p.599) was used vis-à-vis all the theoretical, methodological and the findings/empirical aspects of the sources. The framework sets out 32 research quality evaluation concepts hierarchically interconnected in three tiers through 4 branches, i.e., credibility, contribution, communicability, and conformity. The papers were evaluated against the 23 evaluation concepts at the tail of the concept tree in the framework (See 'Table b'). Since the framework has been developed for the evaluation of research quality including research approach and process in general rather than research output in particular, not all constructing concepts of the framework were strictly applied for this review. The parameters highlighted in Table b were used as a main in/exclusion criteria. In addition, given the literature review was done by one person, attempts were made to reduce subjectivity as much as possible (Kolbe & Burnett, 1991) by relying on the pre-set evaluation criteria to avoid the influence of personal connection or prior knowledge of authors, personal emotions and imaginations.

| | | |
|---|---|---|
| **Credible** | Rigorous | Internally valid |
| | | Reliable |
| | | Contextual |
| | Consistent | |
| | Coherent | |
| | Transparent | |
| **Contributory** | Original | Original idea |
| | | Original procedure |
| | | Original result |
| | Relevance | relevant research idea |
| | | Applicable result |
| | | Current idea (post 1990?) |
| | Generalizable | |
| **Communicable** | Consumable | Structured |
| | | Understandable |
| | | Readable |
| | Accessible | |
| | searchable | |
| **Conforming** | Aligned with regulations | |
| | Ethical | Morally justifiable |
| | | Open |
| | | Equal opportunities |
| | Sustainable | |

**Table b: Literature quality evaluation criteria (adapted from Mårtensson et al., 2016)**

## 3.   Coding and analysis

The next stage of the review process following literature selection was coding and thematic analysis. Prior to abductive review of each paper under each theme (Reichertz, 2007) as per the questions set out in 'Table c', generic themes and topics were identified using a hybrid of manual and software-assisted automated inductive coding. The first set of themes and topics were manually identified by reviewing 117 papers. These papers were selected from the initial relevant literature retrieved by searching databases using the combination of keywords that represent all key concepts, i.e., digital technology/ICT, migration, inequality, and development simultaneously. As elaborated above, however, the keyword search strategy was expanded further and additional literature was retrieved. As the themes started to recur, inspected automatic coding was considered for the additional literature. Consequently, all of the selected publications including those manually thematised were imported to 'NVivo 12 plus' content management and analysis software package and coded inductively. After a careful manual review and validation of the automated coding to address limitations associated with algorithmic evaluation, themes emerged from the manual analysis and from the automated coding were compared, collated, merged, and refined resulting the initial broad themes (See Section 3 and 4 of the main paper). Although incapable of accomplishing thematic analysis by itself, 'NVivo 12 plus' software package has supported the process by offering features that enhance organisation, consistency and automatic coding (Braun & Clarke, 2006). Following the initial thematic analysis, all the selected publications were moved to a separate folder in Mendeley reference manager and tagged according to the themes identified and additional themes were incorporated as they emerged. For consistency and smooth referencing in the future, terms and phrases used for tagging were progressively recorded in a separate document.

| |
|---|
| What aspect of Digital technology, migration, inequality and development the publication primarily focusses on? |
| What are the outstanding concepts and themes in relation to Digital technology, migration, inequality and development discussed in the publication? |
| What is the main theory or argument put forward? |
| What analytical framework if used? |
| What is the key finding? |
| How the argument or finding confirms or contradicts with others? |
| What uniqueness is observed in the study? |
| What methodological or theoretical strength is reflected in the study? |
| Are there any issues or concerns to be noted in the study? |
| What future research gap is identified by the publication? |

**Table c: Guiding question for reviewing publications in various themes**

## 4. Challenges and limitations
### 4.1. Theoretical challenge

The broad, multidimensional, and fluid nature of the concepts of ICTs, migration, and inequality/development has posed theoretical complexity in marrying theories emerged from different dimensions of the literature. Due to the novelty of the study of digital technology in the context of migration, and inequality/development as a nexus, matured and tested analytical frameworks are hardly available notwithstanding this might be considered not only as a challenge but also as an opportunity. Furthermore, use of different nomenclature of concepts such as ICTs and different types of migrants in different publication outlets made it difficult to readily locate relevant publications.

### 4.2. Methodological challenge

The review has been confronted with some methodological challenges. For example, reconciling contents of publication in hard copy format such as paper printed books with the digitally-available publication content has been a challenge. In addition, due to a relatively large number and word count of a range of publications and the availability of only one reviewer, the evaluation, classification, and codding process has been extremely tedious and time consuming. The inclusion of publications written only in English language in the review process can also create a loophole in the review outcome. The review might be also prone to bias as publications were reviewed by one person for in/exclusion decisions.

# APPENDIX II: DIGITAL MIGRATION STUDY

Literature shows that digital methods for migration study are becoming popular (e.g., Leurs & Prabhakar, 2018; Leurs & Smets, 2018; Lu et al., 2016; Messias et al., 2016; Sánchez-Querubín & Rogers, 2018; Schrooten, 2012; Zagheni et al., 2014). Mainly influenced by the postcolonial digital humanities and media and migration studies (Risam, 2019), literature in this arena can be viewed in two broad lenses; digital influence in theorisation—ICTs' influence in re-engineering theories and concepts of migration; and digital influence in methodology—ICTs' influence in the way migration study is conducted.

Regarding the former, literature suggests that the academic community recognises that theorising migration has been messy and problematic. One of the major problems of theorising migration has been the failure to recognise or to adequately reflect the fact that the migration world is a circulatory, dynamic, and complex system rather than organised and stable structure that can be studied with stable methods and theories (Castles, 2010; Diminescu, 2008; Urry, 2012 ). Most migration theories tend to suffer from separating mobilities of migrants from that of sedentary populations, migration routes from metropolitan journeys, cross-country circulations from vicinity movements, regional policies and culture from global influences, etc (Diminescu, 2008). However, there is increasing recognition of the flaws of early migration theories primarily those of the pre-1980s which tended to conceptualise migration in a simple pull-push model by emphasising merely individuals; and lacking synthesis of ethnographic descriptions of migration linking the macro level models of economics, geography and demography (Bakewell, 2008; Castles, 2010; Diminescu, 2008; Faulkner et al., 2019; Nieuwenhuys & Pecoud, 2007). Such conceptualisation of migration in terms of a simplistic push and pull model has been challenged and attracted responses such as Appadurai's (1990) multidimensional model of disjuncture theory (Faulkner et al., 2019).

As Castles (2010) argues, to fit its purpose, the contemporary digital technology-mediated migration needs to be theorised vis-à-vis the "social transformative" perspective at various social, temporal and spatial levels (p. 1583). Local, regional or global migration phenomenon cannot be fully understood without thoughtful consideration of the connectivity among various localities and mediations at different levels. As Messias et al., (2016) agree, despite the increasing expansion of digital technologies, linking theory with data remains a challenge in migration study. Although modern technological developments have improved the collection and use of migration data, matters such as lack of data and limitations related to the production, harmonisation and use of cross-country data particularly from the "Global South" continues to constrain not only the empirical evidence but also the development and testing of migration theories (Messias et al., 2016). For example, the lack and insufficiency of empirical evidence on "South-South" migration remains a challenge. This has posed a limitation both in understanding the contextual dynamics of migration in the region, and in informing the overarching migration theories that can enhance the development of migration theories by delivering evidence for theoretical continuity across "South-North" (Nawyn, 2016). ICTs' increasing entwinement with migration warrants reimagining migration theorisation for a more nuanced understanding of the contemporary ICT-influenced migration.

Regarding methods, the literature indicates that digital technology and online communications highly influence the methodology of the contemporary migration study (e.g., see Dekker & Engbersen, 2014; Kotyrlo, 2019; Leurs & Smets, 2018; Table 1 in the main text). Consequently, various scholars suggest a range of approaches to studying migration in the age of digital technology. However, there is tension between different scholars' approaches which appear to focus on the way the interplay between 'continuity' and 'rupture' associated with migrants and their relationship with their environment across different space and time is explained. For example, in early 1990s, researchers such as Tarrius (1989) cited in (Diminescu, 2008) suggested a simultaneous three layers of analysis to understand the time- and space-based relationships of migrants within their networks and their environment—the "mobility paradigm", namely: movement of proximity, movement within the host space, and large-scale international migrations (Diminescu, 2008). By adding the fourth layer reflecting the role of digital technology in the migration dynamics to Tarrius's (1989) model, Diminescu (2008) suggests that the contemporary digital technology-influenced migration can be better explained by understanding mobility in all its physical, imaginary, and virtual modes, and perceiving the continuum in space and time associated with the multiple movements that are accumulated and articulated in the practices of people.

One application of digital technology in migration research is the use of online data and communication traces. For example, with its limitations, Messias et al. (2016) used "places lived" data from Google+ online data to study migration flows among clusters of countries. As the authors already acknowledged, such application of online data to study complex migration phenomenon such as demography or migration flow can be delicate. The obvious drawback is that such data is not representative of the broader migratory population but the thin slice of it—those who access and use technology. As elucidated in Section 6, the availability and use of a vast amount of online data—often termed as "big data"—as a result of intensive use of ICTs by migrants raises important ethical issues; and rigorous research on this delicate area is deficient (Taylor, 2016).

Generally, ICTs are increasingly influencing the way migration is conceptualised and migration studies are conducted. However, focused, and rigorous research investigating the safe use of digital methods for migration studies is rare.